# Highly curved image sensors: a practical approach for improved optical performance

Brian Guenter[1], Neel Joshi[1]\*, Richard Stoakley[1], Andrew Keefe[2], Kevin Geary[2], Ryan Freeman[2], Jake Hundley[2], Pamela Patterson[2], David Hammon[2], Guillermo Herrera[2], Elena Sherman[2], Andrew Nowak[2], Randall Schubert[2], Peter Brewer[2], Louis Yang[2], Russell Mott[2], and Geoff McKnight[2]

[1] *Microsoft Corporation, One Microsoft Way, Redmond, WA 98052*
[2] *HRL Laboratories, LLC 3011 Malibu Canyon Road, Malibu, CA 90265*
*\*neel@microsoft.com*

**Abstract:** The significant optical and size benefits of using a curved focal surface for imaging systems have been well studied yet never brought to market for lack of a high-quality, mass-producible, curved image sensor. In this work we demonstrate that commercial silicon CMOS image sensors can be thinned and formed into accurate, highly curved optical surfaces with undiminished functionality. Our key development is a pneumatic forming process that avoids rigid mechanical constraints and suppresses wrinkling instabilities. A combination of forming-mold design, pressure membrane elastic properties, and controlled friction forces enables us to gradually contact the die at the corners and smoothly press the sensor into a spherical shape. Allowing the die to slide into the concave target shape enables a threefold increase in the spherical curvature over prior approaches having mechanical constraints that resist deformation, and create a high-stress, stretch-dominated state. Our process creates a bridge between the high precision and low-cost but planar CMOS process, and ideal non-planar component shapes such as spherical imagers for improved optical systems. We demonstrate these curved sensors in prototype cameras with custom lenses, measuring exceptional resolution of 3220 line-widths per picture height at an aperture of *f*/1.2 and nearly 100% relative illumination across the field. Though we use a 1/2.3" format image sensor in this report, we also show this process is generally compatible with many state of the art imaging sensor formats. By example, we report photogrammetry test data for an APS-C sized silicon die formed to a 30° subtended spherical angle. These gains in sharpness and relative illumination enable a new generation of ultra-high performance, manufacturable, digital imaging systems for scientific, industrial, and artistic use.

## 1. Introduction

Nearly two hundred years ago Joseph Petzval showed that thick-lens optical systems focus an object plane onto a curved image surface, whose curvature is determined by the geometry and refractive indices of the optical elements; his field-flattening lens designs became a foundation of modern imaging optics. Lacking a practical curved imaging surface, most imaging system designs are optimized to create a flat focal plane to match a planar digital image sensor or film. In modern lens design, multi-objective optimization tools create optical designs that combine numerous elements to minimize optical aberrations. Extending lens design optimization to include an arbitrarily curved sensor broadens the set of high quality lens designs. Several recent papers have explored the potentially dramatic performance improvements possible with curved sensor surfaces [1-6] including improvements such as 7x reductions in length, 37x in weight, and significantly better modulation transfer function (MTF) and relative illumination, especially at the edges of the image field; the benefits and simplicity of this are evidenced in the human eye [Fig. 1(A)]. These findings demonstrate the profound influence of the flat field requirement on lens design. In current design, following Petzval, field curvature is minimized using multiple elements with opposing field curvature contributions. However, increasing lens elements adds to mass and volume and generally introduces additional contributions to other aberrations that degrade lens performance, especially off axis. The degree of curvature required to obtain significant benefits are particular to the specific optical requirements (field of view, aperture, chromatic bandwidth, etc.). For example in the monocentric lens design form, which uses spherical elements arranged concentrically around a central point, high performance designs are achieved with as little as 2 elements. In this case, from symmetry considerations the subtended sensor curvature required is identical to the field of view. While this form is an idealized use of spherically curved sensors, similar to the Schmidt camera telescope, prior research along with our own lens design studies demonstrate how subtended spherical sensor curvature less than the field of view also permits previously unachievable performance with fewer element than current designs.

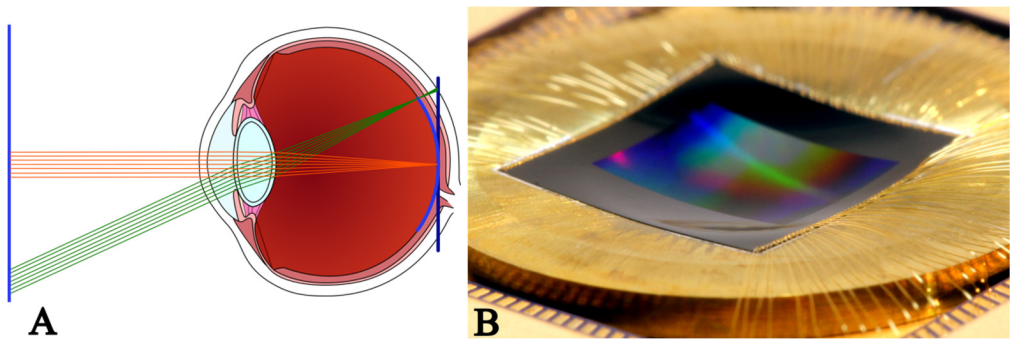

Fig. 1. A) The eye focuses onto photosensitive cells arranged along the curved focal surface inherent of a thick lens. Typical optical lenses require more elements and complexity to focus on flat focal planes and correct the aberrations these compensating elements introduce, losing performance compared to a curved focal plane. B) Functional, 18 megapixel (1/2.3" 7.6 mm x 7.7 mm die) BSI CMOS curved image sensor bonded to a precise 18.74 mm curved mold surface.

Since the lens elements required for a curved field lens are similar to existing flat field lens elements, the challenge to realizing the benefits of curved sensors is creating a highly-curved, high-resolution image sensor. Fabricating an image sensor directly onto a curved substrate would require lithography tools for curved surfaces, a capability which is not yet commercially available [7]. The most investigated methods for creating curved sensors pattern electronics onto flat, flexible substrates that can subsequently be deformed into a curved shape [8-15]. While successful in concept, these approaches have low pixel density due to substrate stretching, incompatibility with sophisticated active pixel CMOS or CCD arrays, and the inability to scale to small pixel pitches (<10 μm) due to processing limitations. An alternative approach is to use a fiber optic faceplate or relay lens to transfer the curved image surface to an arrays of flat sensors [6, 16-17]. This approach adds volume, cost, and complexity and works best with pixel sizes larger than the optical fiber diameter (~2 μm is the minimum); whereas state of the art compact image sensors may have a 1.25 μm pixel size and pitch.

To avoid these limitations, we have developed a precision method for shaping thinned, commercially-available CMOS sensors into highly curved focal surfaces [Fig. 1(B)]. In our method the edges of the die are free to move, which maximizes hoop compressive strains instead of radial tensile strains to increase ultimate curvature. We performed an extensive set of bending experiments on pure silicon samples of many different sizes and shapes ranging from 1/6" to APS-C size sensors. We used these results to validate a numerical model which can predict maximum achievable curvature for any given sensor. Using this model we have simulated different sensor sizes which indicated that the process is largely scale independent and thus widely applicable to various semiconductor sensor types and aspect ratios. To demonstrate the practicality of the method and the performance improvements that are possible we built four identical prototype cameras using spherically curved sensors matching the field curvature of custom $f/1.2$ lenses. We compare our best curved sensor camera with a Canon 1DS Mark III with a 50 mm $f/1.2$ lens and our unmodified flat sensor with a commercial $f/1.2$ lens in Section 3.

## 2. Creating sensor curvature

The single-crystal silicon used for CMOS sensor dies exhibits a combination of high elastic stiffness, low fracture toughness, and high strength (2-5 GPa) that is largely determined by the size and density of surface flaws introduced during manufacturing [18-19]. For single bending axis cylindrical curvature, sensor strains can be minimized according to $e_{xx}=-y/R$ where y is the distance from the neutral axis and R is the radius of curvature, permitting a 25 μm thin die to be curved to a radius <1 cm with stresses well below fracture [20]. However, curving sensors into spherical shapes with two axes of curvature increases stresses due to imposed membrane strains (uniform through thickness) required to accommodate the area change. These stresses cannot be migrated by die thinning.

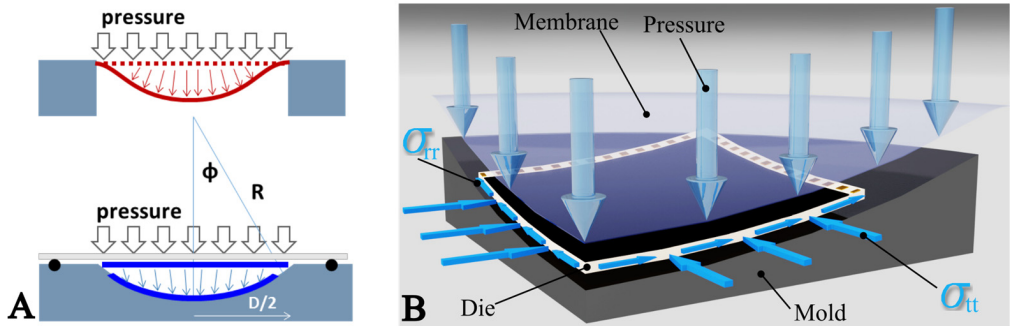

Fig. 2. A) In conventional micro-device 3D membrane formation (red) edges are fixed while pressure or vacuum is applied and deformation is resisted by radial in-plane tensile loads that grow nonlinearly with increasing deflection. By comparison, our pneumatic forming process forms an unconstrained die (blue) through a pressurized flexible membrane that leaves the edges free to translate, thus largely eliminating radial tensile forces and reducing the strain energy density. B) Curving a free standing die into a concave mold introduces in-plane transverse (hoop) compressive strains that grow with increasing curvature. The resulting unstable buckling behavior is prevented through the normal forces applied though the membrane.

Radial and tangential membrane strains that grow nonlinearly with subtended angle occur during mapping of any flat sheet onto a hemisphere. The magnitude and direction of these strains depend solely on the boundary conditions of the deformation process, the shape of the sheet, and the subtended spherical angle. Given the finite strength of silicon, these strains limit the permissible curvature that can be imposed on a sensor die. The non-uniqueness of the deformation mapping raises the question of which forming method maximizes sensor curvature. The bounds for any two dimensional axis of curvature forming method can be established by considering two simple axisymmetric models of a spherical wrapping process [Fig. 2(A)]. In the first, with a fixed perimeter and direct vertical mapping of points from the flat surface to the sphere, the average radial strains are $e_{rr}=(\phi-\sin(\phi))/\phi$ [21] with zero tangential strains, where $2\phi$ is the subtended spherical angle. Conversely, in a second method we imagine mapping the die onto the sphere by wrapping the points around the sphere to create zero radial strains and average tangential compressive hoop strains of $e_{tt}=-(\phi-\sin(\phi))/\phi$ as the die perimeter is forced into smaller radii. The bounds reveal the critical result that membrane strains are both size independent and relate only to the subtended spherical angle. To calculate total strain, we must also add bending strain contributions that are superimposed with the membrane strains. These contributions can be mitigated by creating very thin die with large aspect ratios, but at the risk of increasing instability when subject to compression. This limitation has led other researches to use tensile dominated processes or die with smaller aspect ratios to stabilize the die at the cost of additional bending strains that limit curvature.

In this work, we demonstrate a process that mitigates instability while accruing the advantages of using large in-plane compression to maximize curvature in concave surfaces. In materials with low fracture toughness, such as silicon, compressive strength significantly exceeds tensile strength allowing for greater curvature if peak stresses can be shifted to compression rather than tension. We achieve an approximation of the ideal compressive wrapping by placing a thinned, released die into a forming mold and applying pneumatic pressure to a compliant polymer membrane that deforms and contacts the die progressively from the corners inward as pressure is increased [Fig. 2(B)].

As the die center deflects downward, the die edge is free to move inward, effectively releasing the radial tension but simultaneously building tangential (hoop) compressive stress. Bending moments are created along the perimeter that progressively curve the die into the precise mold shape. As discussed above, the primary difficulty in using free edge forming with thin unsupported plates is suppressing out-of-plane deformations caused by in-plane compression of the thin dies. These instabilities are driven by the same forces that induce wrinkling instabilities in sheet metal drawing processes. The highest risk of instability in our process occurs near the initial contact of the pressure membrane and the die. Pressure on the die surface is reacted by the corners in contact with the mold but not along the die edge, where the middle remains unsupported. A localized, central kink may form when the thickness-to-length ratio if too large. The kink is elastic and can be reversed by removing pressure but will cause rapid failure if increased. One key innovation in this work was the optimization of the mold surface using a conical annulus around the die perimeter that supports the entire edge during initial loading. Thus the mold has a curvature that transitions from spherical near the center to more conical near the die edge. As the shaping progresses, normal loads applied through the pressurized membrane further suppress out-of-plane deformations [Fig. 2(B)].

Finite element simulations were performed with ABAQUS Standard to estimate the strains during the bending process for different curvature states and die thickness. The model was composed of three components: the rigid mold surface, a shell element representation of the silicon die, and a membrane representation of the flexible pressure bladder. The die was discretized into approximately 1000 elements. To simplify the model, we approximated the highly anisotropic silicon as an isotropic material with a Young's modulus of 150 GPa and a Poisson ratio of 0.17. These stiffness estimates average the orientation dependent elastic behavior in the (100) plane between the [100] and [110] directions to provide accurate strain predictions and good approximations of the induced stresses. We assigned a friction coefficient of 0.005 between all contact surfaces. A uniform pressure was applied to the membrane using a smooth step amplitude dependence. We contrast this with a model of the commonly used membrane microfabrication method where the die is selectively thinned to form a membrane while retaining a full thickness wafer surround and then pressurized to deform the thinned area.

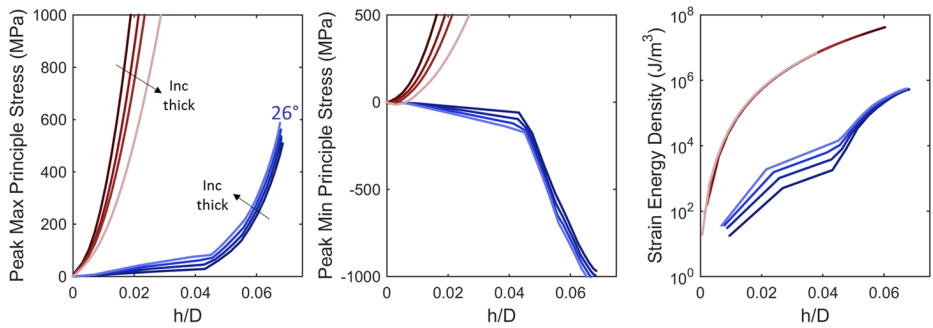

Fig. 3. Finite element predictions for the peak stresses and volumetric strain energy density developed as a function of normalized center displacement for the fixed perimeter and free edge molding operations for thicknesses ranging from 40 μm to 100 μm and a die size of 22.3 mm x 28 mm.

Fig. 3 explores the stress evolution and strain energy density of the free edge and fixed cases to provide insight into their deformation mechanics. For these simulations the specific case of a 22.3 mm x 28 mm die was considered, but the stresses are representative of any size and aspect ratio. Thickness was varied between 40 μm and 100 μm in 20 μm steps. The results are plotted during deformation as a function of center deflection, h, normalized by the die diagonal, D. For the free edge case, we present the results of a process with a final subtended angle of 26° as defined by the rigid mold. Looking at the stress evolution as a function of increasing deflection, the fixed edge case demonstrates the classic plate behavior of transition from bending strains to membrane stretching strains at deflection greater than 3x the thickness. In the case of the free edge forming process, the initial deflections are linked to gross bending of the sheet stabilized by the normal membrane pressure. The sharp increase in stress evolution at h/D = 0.04 coincides with the seating of the perimeter of the die onto the mold surface. After this point, the compressive hoop stresses begin to develop around the perimeter of the die. As the edge becomes increasingly pressed to the mold surface, biaxial tensile strains grow in the center of the die. These results are similar to the stress evolution of a plate subject to pure moment at its edge [23]. Comparing the strain energy density, the fixed edge case builds stored energy quickly and is largely independent of thickness because of the dominance of in-plane tensile stretching. In comparison, the free edge process stores 20-50 times lower energy at a given deflection, and is a stronger function of thickness, especially in the initial stages where bending strains dominate. Near the completion of the forming process, the free edge elastic energy is ~5.5E5 $J/m^3$ which corresponds to an average stress of 405 MPa.

In Fig. 4(A) we present a sequence of FEA model and photogrammetry measurement images of the evolving radial and tangential strains on the die surface of a larger, silicon sample (28 x 22.3 mm, 60 μm thick) as it is formed to a 60.45 mm radius of curvature (30° subtended angle). The images show the large tangential compressive stress that developed at the edge of the die near the center of each of the four edges. This can lead to out-of-plane wrinkling if not sufficiently suppressed by normal force of the pressurized bladder. Using the model with a variety of parameters, we predicted unstable combinations of die aspect ratio and mold curvature that agreed favorably with experimental observations. In experiments, we observed one or more edge wrinkles located near the center of the longest die edge that were generally reversible at small amplitudes but led to rapid failure with increasing pressures.

In Fig. 4(B) we compare the finite element models and digital image correlation experiments. The excellent correlation between the predicted and measured strains suggest that our model is reasonably accurate and thus can be used to predict the stresses for a variety of target shapes and die sizes. Comparing the radial $e_{rr}$ and tangential $e_{tt}$ strains, we confirm that for this free edge forming process, much of the curvature is achieved through tangential edge compressive strains with peaks that are approximately 2x higher in magnitude (-0.7%) than the maximum tensile strains (0.3%). For this particular die and curvature, predicted peak radial strains are located at die center and reach 0.28% and 0.4% for the top and bottom surfaces, respectively.

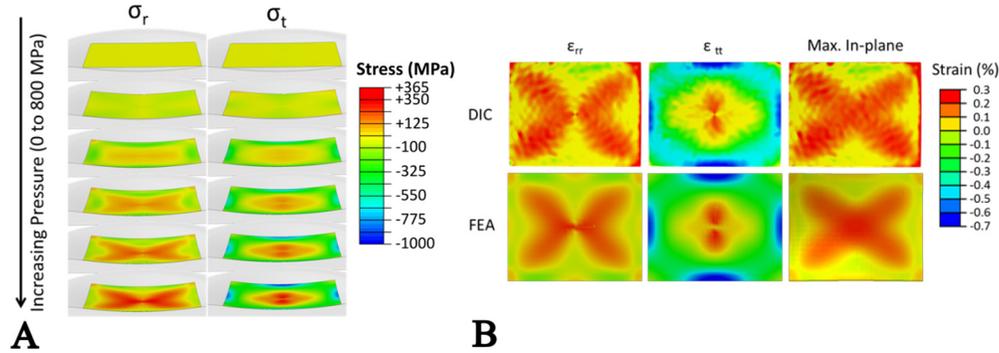

Fig. 4. A) Predicted evolution of radial and transverse stress on the surface of a die as it is forced into a spherical curvature with increasing pneumatic pressure up to 800 MPa. Near the edge, the deformation pattern is compressive in the loop and radial directions, transitioning to biaxial tension near the center. B) Comparison of strains measured using digital image correlation (DIC) on the top surface of the die (upper row, see **Visualization 1** for a time-lapse video) and FEA model estimates (lower row) reveal purely tensile strains along the radial direction, and larger tangential strains that shift from large compression near the edges to tension near the center. Close correspondence of the measured and model results illustrates that our FEM models are a good predictor of the strains in bending actual CMOS die.

Referring again to the two bounding membrane behaviors of pure radial stretching and tangential stretching, this bending method evidently produces elements similar to tangential compression near the edges but radial stretching near the center. This stress profile is qualitatively similar to analytical models for nonlinear deformation of circular plates curved purely by edge moments [23]. Based on our deformation technique, this analogy should provide similar deformation mechanics. Under the free edge boundary conditions, an inward bending moment is created due to the interaction of the rigid mold and the suspended die pushed upon by the pneumatic pressure. This moment continually progresses inward as the die is formed.

While strain energy density is minimized using our approach, considerable elastic energy is stored within the highly thinned die as curvature is increased. Depending on die thickness and curvature, we form dies with pressures varying between 6 and 15 atm and we estimate stresses in the dies can exceed 1 GPa in compression and 500 MPa biaxial tension in the die center. We found that despite this large stored energy, a thin (~1 μm) layer of low viscosity epoxy adhesive was sufficient to bond the die to the mold surface and hold a precise shape tolerance. We temperature cycled a bonded sensor between -20° C and 50° C and did not observe shape changes.

The free edge deformation profile is advantageous from the standpoint of image sensors, since strains are smaller in the photo-sensor region located near the die center than in the peripheral used for readout and control electronics. Limiting strains in the photosensitive central region of the die reduces any potential detrimental effects of mechanical strain on electro-optic properties including dark current and quantum efficiency.

The curvature limits for this process are defined by the failure strength of the silicon and processed CMOS die along with optimization of mold shape and friction coefficients. In our experiments, we observed failure in silicon samples and CMOS devices when the FEA predicted compressive stresses exceeded 1.4 GPa and maximum biaxial tensile stresses exceeded 500 MPa. We were unable to determine definitively if failures initiated in the tensile or compressive region due to the elongated fracture or shattering of die. Assuming linear elastic fracture mechanics and a mode-I critical fracture intensity of 0.9 MPa m$^{1/2}$ along with the measured surface roughness of 80 nm RMS, we estimate a mean failure stress of 1.9 GPa for pure silicon samples. The back-side-illuminated sensor die are composed of a thin silicon absorber mated to a multi-layer CMOS heterogeneous readout structure made up of oxide and metal structures [Fig. 6(A)]. Given this heterogeneous structure along with stress concentrations due to contact with the rigid mold, the observed strengths may be consistent with expected performance. Reducing surface flaw sizes and minimizing contact induced stresses will allow for even greater curvature through higher strain capacities [18].

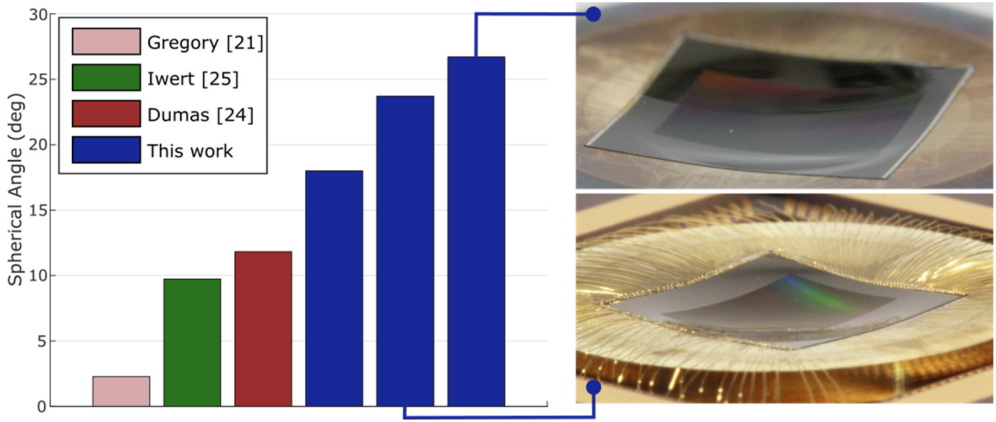

Fig. 5. Comparative graph of curvature achieved in working sensors between this work and the significant work from the literature. A wirebonded sensor used for one camera in this study is shown in the lower right, having a spherical curvature of 23.7°. The working sensor in the upper right has a curvature of 26.7° but could not be used for this study as it does not match the lens.

While the specific sensor curvature required depends on a given lens design, our studies and literature findings indicate that optimal sensor curvature tends to increase as field of view increases. Our current results of 23.7° (the prototype camera sensor curvature) and 26.7° (our maximum to date for a working sensor) are twice the curvature that has been previously reported for working sensors [21, 22, 24-26]. A comparison of the curvatures we have realized with working image sensors is graphed in Fig. 5, with photographs of the curved and bonded sensors at 26.7° and 23.7° shown at the top and bottom right side, respectively. The largest curvature in a bulk silicon die we have achieved is the 30° subtended angle shown in Fig. 4, 150% more than the silicon sample shown by Dumas[24] and with more precise shape accuracy. Gregory [21] demonstrated a 7° field, *f*/1 telescope capable of imaging faint objects over a wide field of view, enabled by a large curved imager mosaic with high surface accuracy (conformance of 1.7 μm across an array 120 x 180 mm curved to 5.44 m radius). Dumas demonstrates essentially equivalent optical performance on-axis between the planar and curved bolometer sensors, but with improved uniformity as the angle of incidence increases in the latter case. The "double-vacuum" method used by Iwert [25] achieves curvature similar to that of Dumas [24] with similarly low shape accuracy, especially in the center. The flat center is likely a result of the vacuum forces being insufficient to completely shape the sensors, something we also observed as we developed our process.

### 3. Prototype camera design and measurements

Using this bending technique, we have deformed stock commercial CMOS die for integration into a prototype camera. The die is a back side illuminated (BSI) 18 megapixel CMOS sensor (Aptina AR1820HS) with a 1.25 μm pixel pitch; the die is 7.6 x 7.7 mm, and the light sensitive area is 6.1 x 4.6 mm. The substrate was thinned to 16 μm, leaving a total die thickness of 25 µm and a length to thickness aspect ratio along the diagonal of 480:1. An SEM image of this sensor's layers is given in Fig. 6(A). We have deformed the sensor portion of these dies into curvatures with radius between 24.6 and 16.7 mm realizing subtended spherical angles between 18.0° and 26.7° along the diagonal. The sensors used in our cameras have an 18.74 mm radius of curvature (23.7°.) The dies were bonded to the mold surface using an epoxy adhesive cured at a temperature of 100° C for 30 minutes; later samples were successfully bonded using solder, reducing the total forming rate to only a few minutes in total. After the adhesive had fully cured, the die was cooled and the pressure released to achieve a free standing die with deviations from the targeted radius of curvature of 0.3 µm RMS and 2.2 µm peak to valley, measured interferometrically [Fig. 6(B)]. Note the depth of focus at $f$/1.2 is approximately 3 µm, and the accuracy of the molds is between 0.56 and 1 µm peak to valley. The sensor and mold together were mounted and wirebonded to a ceramic leadless chip carrier (LCC) package using a manual ball bonding machine. A custom adapter interfaces the LCC to the manufacturer's development kit. We measured the dark current for all our wirebonded sensors, see Fig. 6(C), and observed a slight improvement in dark current over the COTS sensor, within the experimental error of the test.

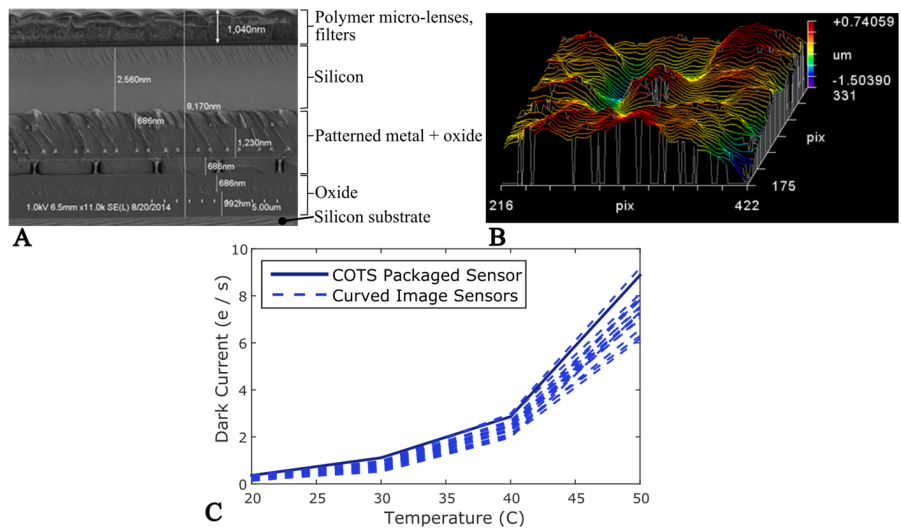

Fig. 6. A) SEM cross-sectional image of the CMOS sensor used in this work, demonstrating the variety and complexity of the materials undergoing deformation. The silicon substrate shown at the bottom has a 16 µm thickness and is outside the field of view. B) Interferometer measurement of the shape accuracy of a camera sensor at 2.2 µm peak-to-valley. C) Plot of the measured dark current versus temperature of all wirebonded sensors curved to the lens prescription. Most of the curved samples have slightly lower dark current than the unmodified flat sensor.

To quantitatively assess the optical performance benefits of a curved sensor, we designed several lenses in both curved and flat form, with fields of view from 30 to 75 degrees. We chose one design to build based on wafer availability (a wafer-level process is essential). For the built system we show both as-designed and measured performance values. For the others we show only as-designed performance.

The built system has an $f$/1.2 lens with a 55-degree field of view for a 7.6 mm image circle. The lens has 8 glass elements with 5 aspherical surfaces and is optimized for visible wavelengths from 450 to 650 nm. Fig. 7 shows the as-designed MTF of the curved and flat sensor systems. The curved system has almost double the sharpness at every field angle, despite having 1 less element and 2 fewer aspherical surfaces.

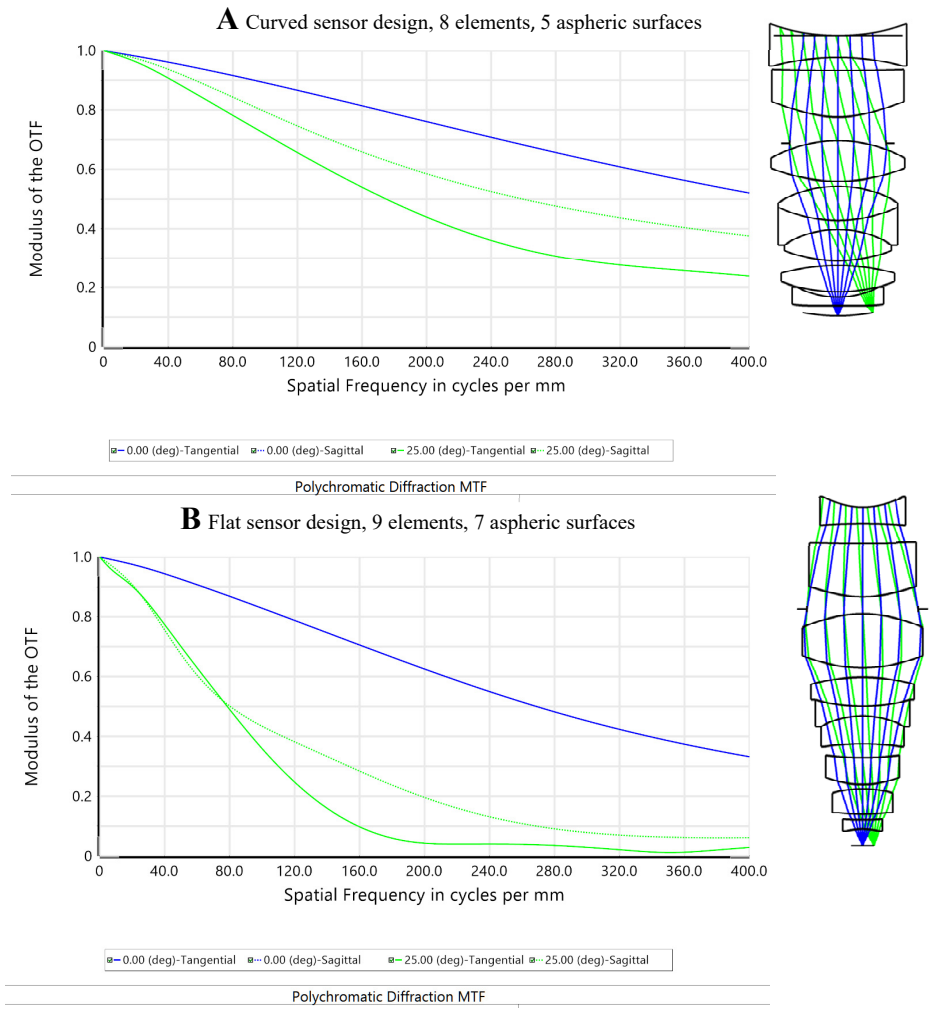

Fig. 7. As-designed MTF for curved vs. flat designs for $f/1.2$, 1/2.3" CMOS sensor with 50 ° field of view. A) Curved sensor, optimized for an 18.74 mm radius of curvature sensor. B) Flat sensor optimized design. The curved system has almost double the sharpness at every field angle, despite having 1 less element and 2 fewer aspherical surfaces. The flat sensor design is also much larger.

Fig. 8 shows the measured performance of two flat-field $f/1.2$ lenses with similar field of view vs. the measured performance of the curved sensor lens. For the flat field lenses, we chose two high quality commercially available lenses 1) a Canon 50 mm $f/1.2$ lens mounted on a Canon 1DS Mark III professional camera body, and 2) and an Edmund Optics 6 mm $f/1.2$ lens mounted on the stock non-thinned, flat version of the AR1820HS sensor. The results were obtained by measuring all three cameras using the slant edge technique (ISO 12233:2014), as implemented in the industry standard Imatest software, with diffuse front illumination from a white light with a spectrum covering approximately 450 to 650 nm. Comparing the modulus transfer function (MTF) charts of the two lenses one can see that the curved sensor in Fig. 5 has more than double the sharpness (as measured by the MTF30 response) of the conventional flat sensor lenses at any given field angle, despite having fewer and less complex lens elements.

While performance is increased throughout the field, the edge and corner performance is particularly improved compared to the flat baseline, showing close to a 5x improvement at the edge and more than 3x improvement in the corner. Our lens shows exceptional performance for an *f*/1.2 lens. We have designed several other lenses with different focal lengths for curved and flat sensors. These curved sensor lenses are also significantly sharper than the similarly-optimized flat sensor designs. In addition, Fig. 8 shows measurements of the relative illumination across the field of our curved sensor prototype. The image has considerable uncorrected distortion. This was intentional; we did not attempt to minimize distortion in the lens design optimization because it is easily corrected as a digital post-process.

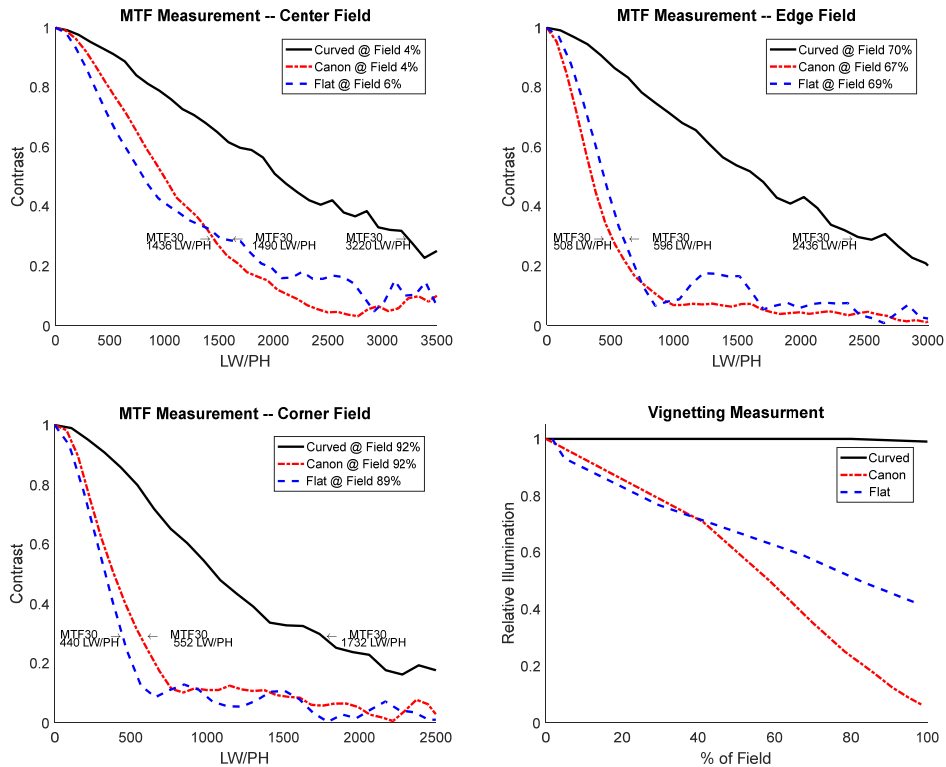

Fig. 8. Measured MTF results and relative illumination performance with comparisons to two high-quality commercial *f*/1.2 cameras. Curved refers to the prototype curved CMOS sensor camera. Canon refers to a Canon 50 mm *f*/1.2 lens mounted on a Canon 1DS Mark III body. Flat refers to an Edmund Optics 6 mm *f*/1.2 lens mounted on a flat version of our CMOS sensor. The data show more than double the sharpness in the center and more than triple in the corner, as measured in Line-Width/Picture-Height (LW/PH) at MTF30. Relative illumination measurements show virtually no light lost from center to corner for our curved sensor prototype compared to over 90% loss for the Canon lens.

Our curved camera prototype maintains near perfect consistency in illumination, losing only 1% of the illumination of the corner relative to the center. It vastly outperforms the Canon lens which diminishes by 93.75% (4-stops) of illumination in the corner, while there is a 1.25 stop improvement over the Edmund Optics lens. Fig. 9 is a representative shallow depth of field image captured using the *f*/1.2 curved sensor camera prototype. At a resolution of 3220 (center) and 1732 (90% field) line widths of resolution per picture height, this camera resolves more detail and has more uniform illumination than a much larger professional 35mm camera system fitted with a *f*/1.2 50mm lens.

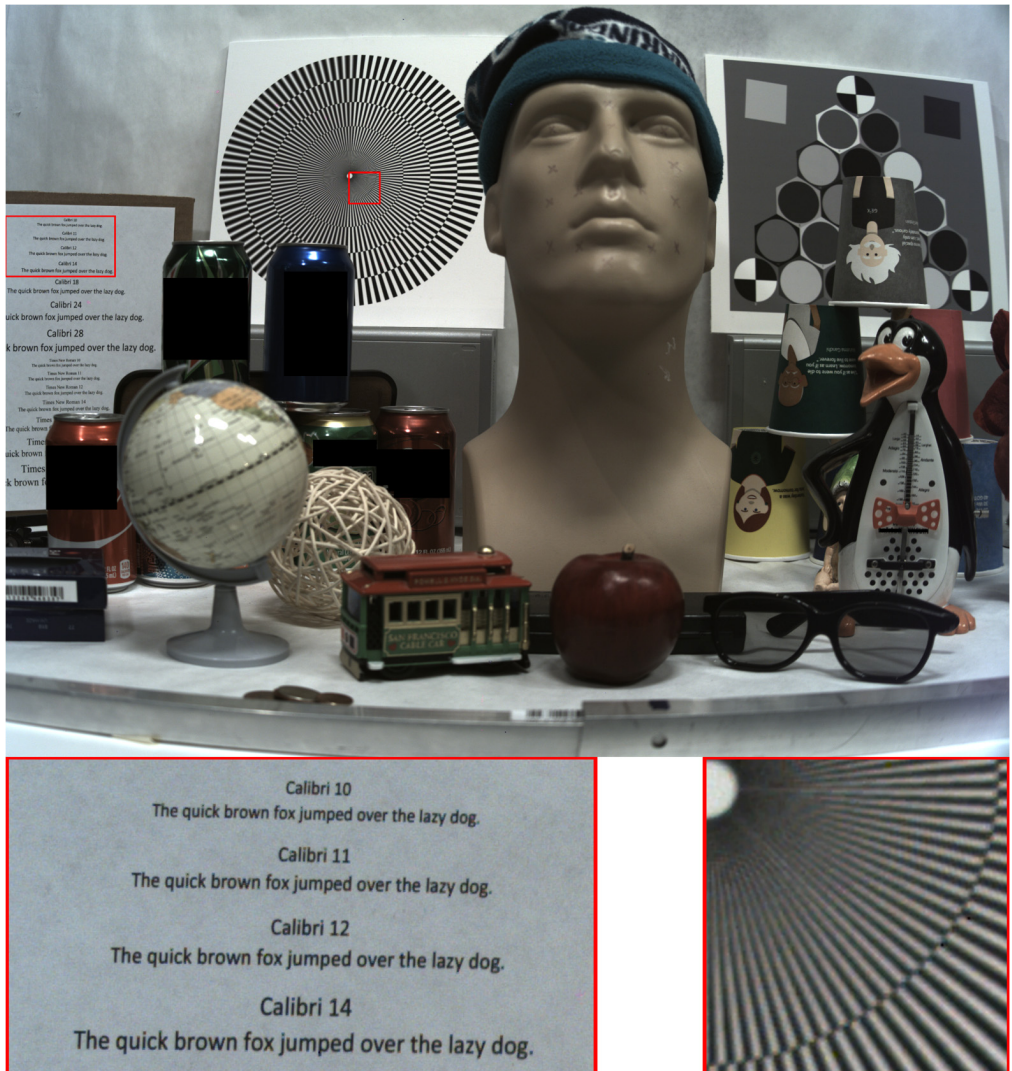

Fig. 9. Image captured with the curved sensor prototype camera. The zoomed-in portion shows excellent sharpness far from center. The large depth range in the scene illustrates pleasing shallow depth of field enabled by the wide aperture of the *f*/1.2 lens.

Fig. 10 and Fig. 11 present two additional lens design studies comparing lens performance for optimized flat and curved sensors covering a 24 x 36 mm sensor while constraining sensor curvature to a maximum of 36° of subtended angle. Fig. 10 shows as-designed performance for a wide angle 28 mm *f*/1.7 lens (using 35.9° subtended angle). The curved design has almost 1.3x the spatial frequency response despite having three fewer elements and one asphere versus two for the flat design. Fig. 11 shows as-designed performance for curved and flat sensor 80 mm *f*/1.5 30° field of view lenses. Again, the curved system (with 29° subtended sensor angle) has significantly better spatial frequency response with five elements than the flat design achieves with eight. We believe that further development could achieve these curvatures with commercial sensors.

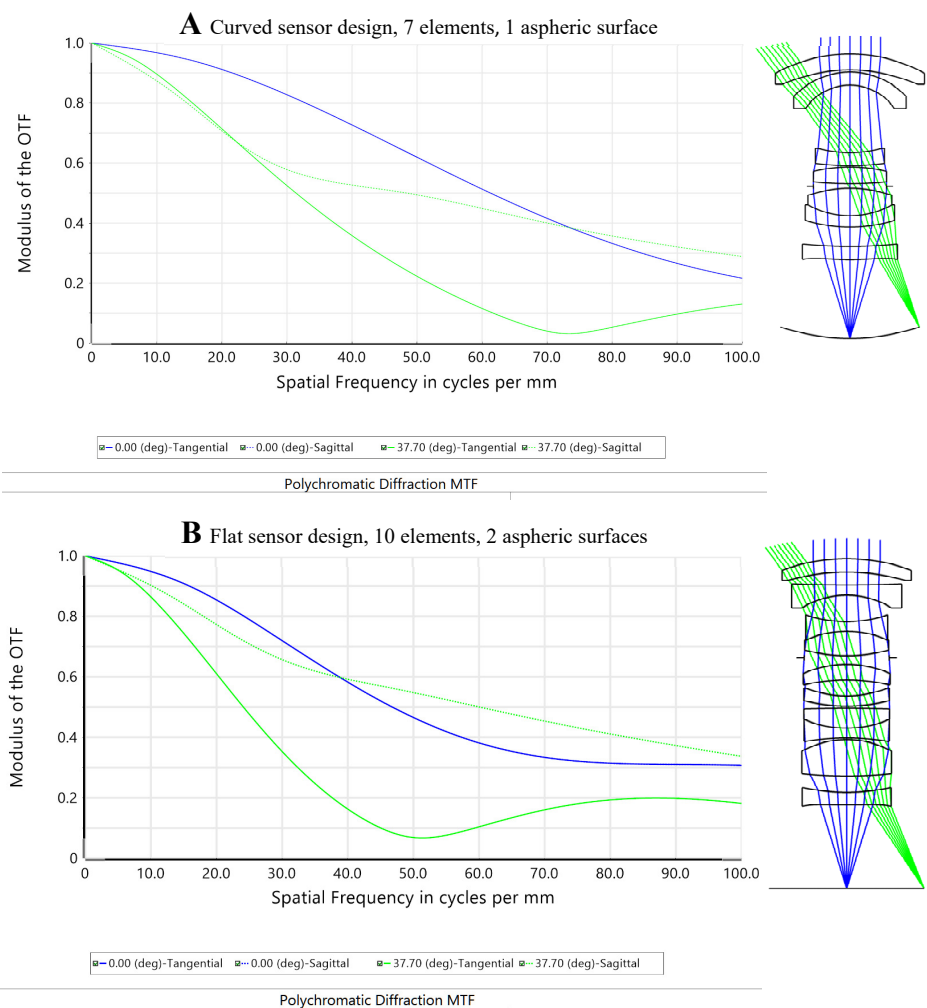

Fig. 10. Design study comparison between the curved sensor optical system A) and flat sensor B). 28 mm focal length, *f*/1.7, 75° diagonal field of view, 43 mm image circle. Flat lens B) has 10 elements and 2 aspherical surfaces. Curved sensor lens A) has seven elements and one aspherical surface. At MTF50 the curved sensor design A) has 1.3x the spatial frequency response of the flat design B) despite having fewer elements.

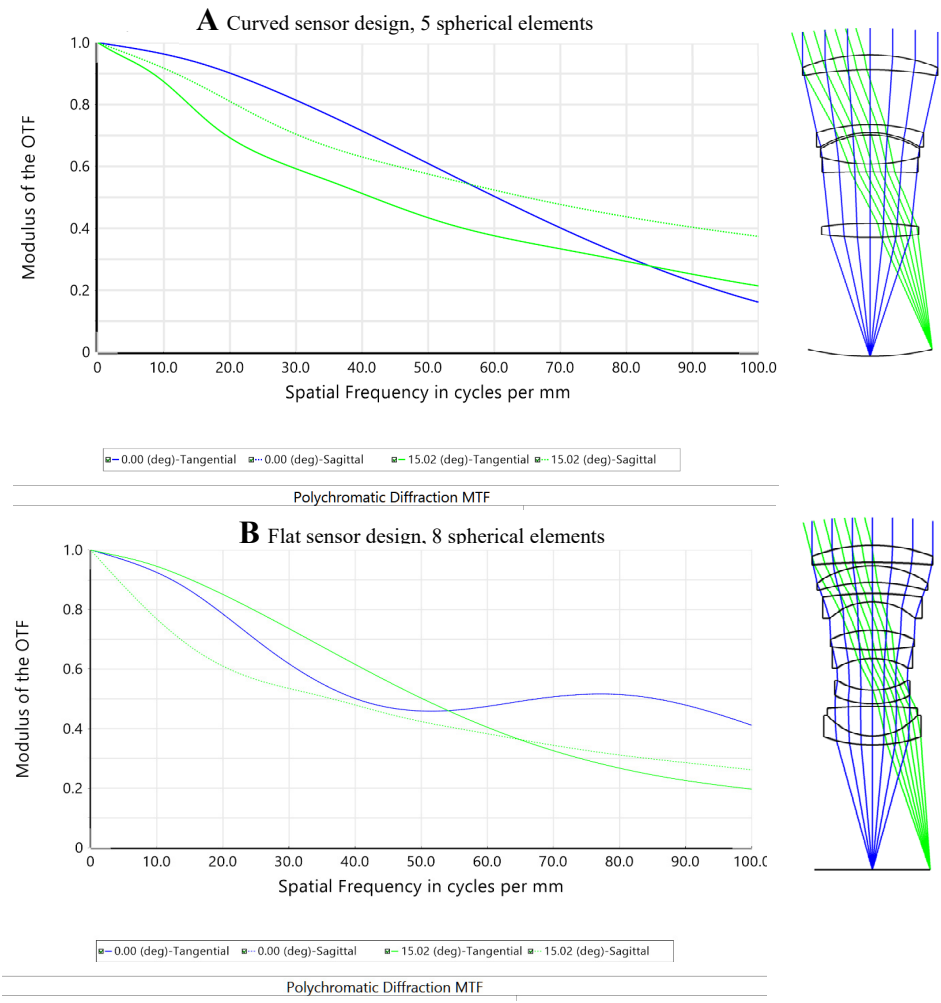

Fig. 11. Design study comparison for 80 mm, $f/1.8$, 30° field of view, 43 mm image circle. Flat sensor lens B) has eight spherical elements. Curved sensor lens A) has five spherical elements. At MTF50 the curved sensor design A) has 1.2x the spatial frequency response of the flat design B) despite having three fewer elements.

## 4. Conclusions

Curved sensor technology is on the cusp of making a large impact on a number of scientific fields including photography, videography, computer vision and automation, reconnaissance and surveillance imaging, microscopes, and telescopes, among others. Curving the image surface can improve performance along many performance axes: illumination uniformity, resolution, light-gathering while also reducing system size, cost, and complexity. Using our optimized free edge bending process, we have created a prototype curved sensor camera with extraordinary performance. It surpasses, and in the case of illumination uniformity far exceeds, the performance of much larger professional camera systems like the Canon 1Ds DSLR equipped with a similarly fast $f/1.2$ 50 mm lens. The flexible bending process is scalable to any die size, optically precise even for fast lenses, and compatible with cost effective, large scale manufacturing processes.